\documentclass[aps,preprint,showpacs,superscriptaddress,groupedaddress]{revtex4}  

\usepackage[utf8]{inputenc}
\usepackage{amsmath}
\usepackage{amssymb}
\usepackage{graphicx}
\usepackage{tikz}

\newcommand{\be}{\begin{equation}}
\newcommand{\ee}{\end{equation}}
\newcommand{\bea}{\begin{eqnarray}}
\newcommand{\eea}{\end{eqnarray}}
\newcommand{\beann}{\begin{eqnarray*}}
\newcommand{\eeann}{\end{eqnarray*}}

\newcommand{\Dt}{\Delta \theta}

\DeclareMathOperator{\Tr}{Tr}
\DeclareMathOperator{\dd}{d}

\begin{document}

	\title{Localization of gauge fields and Maxwell-Chern-Simons theory}

	\author{Daniel Flassig}
	\email{daniel.flassig@physik.uni-muenchen.de}
	\affiliation{Arnold-Sommerfeld-Center, Ludwig-Maximilians-Universit\"at,\\ 
		         Theresienstr. 37, D-80333 M\"unchen, Germany}
	\affiliation{Physik-Department, Technische Universit\"at M\"unchen, \\
	             James-Franck-Str., 85748 Garching, Germany}
	
	\author{Alexander Pritzel}
	\email{alexander.pritzel@physik.uni-muenchen.de}
	\affiliation{Arnold-Sommerfeld-Center, Ludwig-Maximilians-Universit\"at,\\ 
		         Theresienstr. 37, D-80333 M\"unchen, Germany}
	
	\preprint{LMU-ASC 43/11}

	\begin{abstract}
		We propose an explicit model, where an axionic domain wall dynamically localizes a U(1)-component of a nonabelian gauge theory living in a 3+1 dimensional bulk. The effective theory on the wall is 2+1d Maxwell-Chern-Simons theory with a compact U(1) gauge group. This setup allows us to understand all key properties of MCS theory in terms of the dynamics of the underlying 3+1 dimensional gauge theory. Our findings can also shed some light on branes in supersymmetric gluodynamics.
	\end{abstract}

	\maketitle

    \section{Introduction}
	Maxwell-Chern-Simons (MCS) Theory in 2+1 dimensions is a theory leading to a plethora of interesting phenomena, it has found a wide array of applications ranging from fractional quantum hall effect \cite{Zhang:1988wy} to knot theory \cite{Witten:1988hf}.
   
   Some of the more interesting implications of adding a Chern-Simons (CS) term include screening of the electric field, the association of electric charges with magnetic fluxes and the appearance fractional statistics for charged particles \cite{Wen:1988cx}. Compact U(1) gauge theories, which ordinarily are confining \cite{Polyakov:1976fu} in 2+1 dimensions are altered in an even more dramatic way. The confinement disappears \cite{Kogan:1995ei} and magnetic events (the analog of sphaleron transitions) start producing electrically charged particles \cite{Bachas:2009ve}. Also, the coefficient of the CS term can only take discrete values \cite{Deser:1982vy}.
   
   In this letter, we are going to propose an explicit setup in the Dvali-Shifman gauge field localization framework \cite{Dvali:1996xe}, in which a 2+1 dimensional MCS theory appears as the effective world volume theory of a field theoretic brane in a 3+1d spacetime. This is achieved by promoting the field theoretic brane into an axionic domain wall. We can then clarify many of the interesting features of 2+1d MCS theory in terms of 3+1d bulk physics. 
   
  \section{2+1d Chern Simons Electrodynamics}
  We begin by summarizing various properties found in 2+1d Maxwell-Chern-Simons theory with a $U(1)_\mathrm{c}$ gauge group. Its action is given by
\be
 \label{CS term}
 S=\int_\textrm{2+1d}-\frac{1}{2g^2}F\wedge \star F+\frac{k}{2}A\wedge dA +A \wedge \star j
\ee 
  Here the Chern-Simons term acts as a mass term with mass proportional to $kg^2$ and screens the electric field. Furthermore by explicitly solving the equations of motion for a point charge we can see that a point particle with electric charge $qe$ at $\vec{x}_0$ produces a magnetic field $B$ of
  \be
  	B(\vec{x},t)=\frac{q}{k}e\delta (\vec{x}-\vec{x}_0).
  \ee
    
  A further peculiarity is the appearance of fractional statistics first noticed by Wen and Zee \cite{Wen:1988cx}, i.e. exchange of two particles of electric charges $q_1,q_2$ leads to an additional phase of 
  \be
\label{fractional statistics CS phase}
  	\delta=\frac{1}{2k} q_1 q_2
  \ee
   in the wavefunction.
  
  It was shown by Polyakov \cite{Polyakov:1976fu} that QED with a compact U(1) gauge group in 2+1d is confining at exponentially large scales due to instantons, which effectively act as a plasma of magnetic charges and lead to Debye screening, which in turn implies the existence of a mass gap. However in \cite{Kogan:1995ei} it was noted that confinement is absent if one introduces a CS term.
   
   Gauge invariance under large gauge transformations demands that k must be quantized \cite{Polychronakos:1990xq} as
  \be
  	k=\frac{n}{8\pi}\hspace{1cm}n\in \mathbb{N}
  \ee

   \section{Dvali-Shifman localization mechanism}
   In \cite{Dvali:1996xe} Dvali and Shifman proposed a field theoretic mechanism for localizing 3+1d gauge fields to 2+1d solitons.
Their setup includes a strongly interacting nonabelian gauge theory, e.g. SU(2), living in 3+1d space (bulk) and a Higgs field in the adjoint representation. In the bulk, their Higgs has a vanishing vacuum expectation value (vev) and the gauge theory is in the confining phase. Their explicit theory allows for a domain wall (brane) on which the Higgs acquires a nonzero vev --- the SU(2) is broken down to U(1)$_\mathrm{c}$ on the wall. Because electric flux cannot escape into the bulk, this setup actually produces an effective 2+1 dimensional U(1)$_\mathrm{c}$ gauge theory with compact gauge group on the brane.

Their theory can also be understood as the electric-magnetic dual of a Josephson junction \cite{Dvali:2007nm}, where a layer of insulator is sandwiched between two superconductors. Magnetic flux is confined to the junction while charges inside the junction are still screened. 
   
   This duality lends itself to a straightforward interpretation of confinement in 2+1d QED with compact U(1). In the Dvali-Shifman picture, monopoles from the condensate in the bulk can tunnel across the brane (the dual effect of the Josephson current). The tunneling monopoles effectively act as a dilute plasma of magnetic charges on the brane and induce confinement in the 2+1d theory. 
   
   \section{Induced Chern-Simons Term}
It is well known that the effective world volume theory of an axionic domain wall contains a Chern-Simons term. Starting with a (3+1)d Yang-Mills theory, which contains an axion $\theta(x)$ coupling to the SU(2) gauge fields
\be
	S^\text{(3+1)d}_\theta = - \frac{1}{8 \pi^2} \int \theta\Tr F \wedge F
\ee
This can be written as a total derivative of the Chern-Simons form
\begin{equation}
	S^\text{(3+1)d}_\theta= - \frac{1}{8 \pi^2} \int \theta \dd \Tr [A \wedge \dd A + \frac{2}{3} A \wedge A \wedge A]
\end{equation}
and after integrating by parts gives a contribution on an axionic domain wall, across which the axion VEV changes by $\Delta\theta$,
\begin{equation}
	S^\text{(3+1)d}_\theta= \frac{\Delta\theta}{8 \pi^2} \int_\text{(2+1)d} \Tr [A \wedge \dd A + \frac{2}{3} A \wedge A \wedge A]
\end{equation}

Comparing to (\ref{CS term}), we find the coefficient of the induced CS term to be 
\begin{equation}
	\label{CS coeff}
	k = \frac{\Delta \theta}{8 \pi^2}
\end{equation}

	\section{Localization of Chern-Simons theory}
We will now consider the situation, where a gauge field is localized to an axionic domain wall. For concreteness, consider again a confining SU(2) gauge theory in the bulk, with both an axion field and a Higgs in the adjoint representation. The explicit potential presented below allows for an axionic domain wall and gives a nonzero Higgs-vev on the domain wall only.

Assuming the confinement scale in the bulk to be sufficiently high, there will again be a $\mathrm{U(1)}_\mathrm{c}$ gauge theory localized to the brane. The jump in the axion value across the brane however contributes a CS term to the 2+1d effective action, with a coefficient given by (\ref{CS coeff}). 
	
	\begin{figure}
\setlength{\unitlength}{1pt}
\begin{picture}(226,216)
	\newsavebox{\leftfig}
	\savebox{\leftfig}(108,216)[bl]{
		\put(0,0){\includegraphics{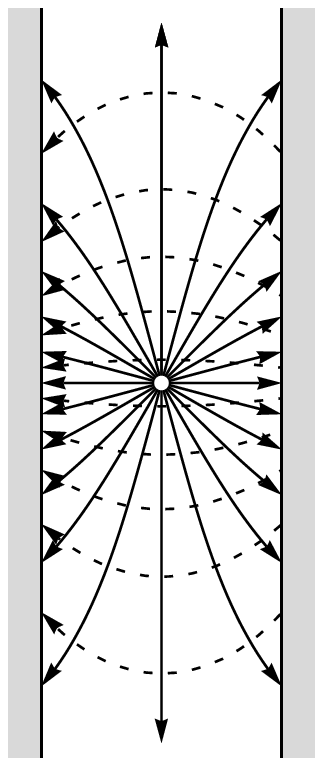}}
	}
	
	\newsavebox{\rightfig}
	\savebox{\rightfig}(108,216)[bl]{
		\put(0,0){\includegraphics{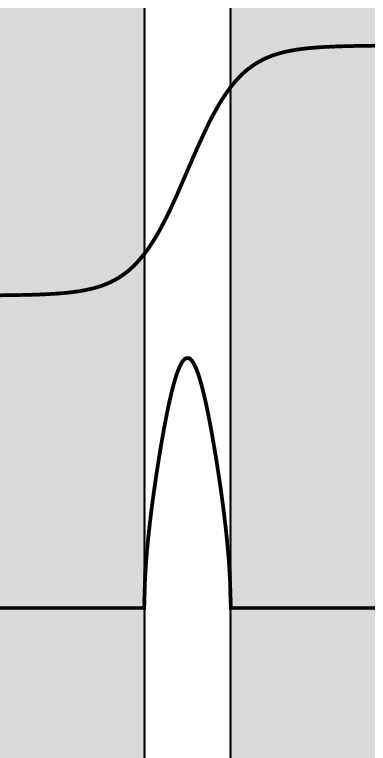}}
		\put(9,14){SU(2)}
		\put(45,14){U(1)}
		\put(76,14){SU(2)}
		\put(9,170){Axion}
		\put(9.5,80){Higgs}
	}
	
	\put(0,0){\usebox{\rightfig}}
	\put(118,0){\usebox{\leftfig}}
\end{picture}
\caption{Axion and Higgs field configuration across the domain wall on the left. And U(1)-electric (solid) and magnetic field lines (dashed) of an electric point charge within the wall on the right.}
\label{fig brane field}
	\end{figure}
	
	We now elaborate on the 2+1d effective field theory that we expect to see in our model. We will then take a complimentary perspective and compare the 2+1d arguments to the 3+1d setup in an accessible limit. Along these lines we will clarify the appearance of a photon mass.
	
	In 2+1 dimensions, pure Maxwell-theory contains a massless photon, which corresponds to one propagating degree of freedom(dof). As the 2+1d theory that we are interested in comes from a 3+1d model, we also expect to see a tower of heavy states. Their characteristic mass scale is given by the inverse localization width and they appear as Proca-fields in the Lagrangian, containing 2 dof.
	
	When a Chern-Simons term is added to the 2+1d Maxwell theory, the previously massless photon acquires a mass $m_\mathrm{cs} = k g^2$. A vector field with a Proca-mass of $m_\mathrm{p}$ on the other hand splits up. One of its degrees of freedom becomes lighter, the other one heavier, with masses
\begin{equation}
	m_\pm = \sqrt{m_\mathrm{p}^2 + \frac{m_\mathrm{cs}^2}{4}} \pm \frac{m_\mathrm{cs}}{2}
\end{equation}	
The number of degrees of freedom per field however does not change in either case.

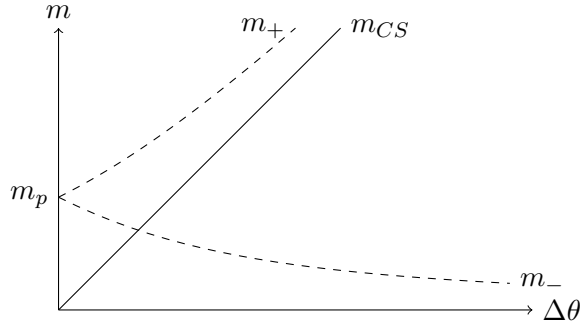
\begin{figure}	
	\begin{tikzpicture}[domain=0:4,scale=1.5]
    		\draw[->] (-0,0) -- (4.2,0) node[right] {$\Dt$};
    		\draw[->] (0,0) -- (0,2.5) node[above] {$m$};
     		\draw[color=black] plot [domain=0:2.5] (\x,\x)
        					node[right] {$m_{CS}$};
        		\draw[style=dashed] plot[domain=0:2.1] (\x, {sqrt(\x^2/4+1)+\x/2})
        					node[left] {$m_+$};
		\draw[style=dashed] plot (\x, {sqrt(\x^2/4+1)-\x/2})
        					node[right] {$m_-$};
        		\draw(0,1) node[left] {$m_p$};
	\end{tikzpicture}
	
        \caption{\label{fig eft masses} 
        	Mass of the zero mode and the first Kaluza-Klein modes as a function of $\Dt$
        }
\end{figure}

	For the effective 2+1d theory, we are interested in the lowest energy degrees of freedom. The field at long distances will be completely dominated by the lightest mode in the spectrum. If the CS term is small, the previously-massless-photon will be this lightest dof with mass $m_\mathrm{cs}$. If we imagine to increase the CS term, there will be a crossover at some point and the $m_{-}$ mode of the first level Proca field becomes the lightest mode, as illustrated in fig. \ref{fig eft masses}. 
	
	Let us now return to our 3+1d model and study the electromagnetic field sourced by a charge on the brane. Consider the brane on which SU(2) is broken to U(1)$_\mathrm{c}$ to be of finite thickness. Further assume a simplified axion field configuration, where the change of the axion vev is localized at the left and right boundary of the brane (unlike the axion field depicted in fig. \ref{fig brane field}, left). Then the U(1) field on the brane worldvolume just obeys Maxwell equations and has some peculiar boundary conditions at the interfaces to the bulk. The static field of a point charge can be calculated along the lines of the image charge techniques developed for topological insulators in \cite{Karch:2009sy}. 
The results are in perfect agreement with the expectations from the lower dimensional effective theory. For a configuration corresponding to the lowest CS-level the electric field $\vec{E}$ decays for large distances like a Yukawa potential, i.e.
	\be
		\vec{E}\sim \frac{1}{\sqrt{r}}e^{-kg^2r},
	\ee
while the image charges produce a localized magnetic flux across the brane as expected. The flux lines corresponding to this configuration are depicted in fig. \ref{fig brane field}, right. Furthermore if we check for more general jumps of $\theta$ across the brane, we see that the effective mass varies as predicted, i.e. it grows linearly until $kg^2\sim1$ and then drops again as $\Dt$ is raised further, reproducing the expected $1/\Dt$ behavior. 
See the appendix for more details. 

	\section{Magnetic events and confinement }
	Let us shortly recall magnetic events and then elaborate on their physics in our gauge field localization setup. In 3+1 dimensions, magnetic monopoles in U(1)$_\mathrm{c}$ gauge theory are well known. These static field configuration in 3 space dimensions however can also be interpreted as instantons of a euclidean 2+1d gauge theory. Such tunneling transitions between topologically inequivalent vacua of the classical gauge theory play a crucial role in Polyakov's derivation of confinement. The associated classical transitions in Minkowski-space were recently dubbed magnetic events. In pure U(1)$_\mathrm{c}$ theory, their field evolution is singular at one spacetime point, yet this singularity is lifted, when the theory emerges from a broken SU(2). Moreover, the events carry magnetic charge. Finally, in the context of MCS theory, consistency demands that an electrically charged particle remains as a ``remnant'' of a magnetic event \cite{Bachas:2009ve}.
	
	In our model, one can understand the magnetic events as actual monopoles flying across the brane. A monopole crossing the brane will acquire an electric charge of 
	\be
		q=\frac{\Dt}{2\pi},
	\ee 
	as shown by Witten \cite{Witten:1979ey}. The electric flux in the bulk however is confined into flux tubes, therefore the monopole will stay connected to the brane by a flux tube. From the lower dimensional point of view the throat of that tube looks like an electric point charge, precisely as seen in \cite{Bachas:2009ve}. Note that this is consistent with conservation of the electric current, as from the (2+1) d effective field theory point of view the magnetic event generates a Chern-Simons current. In the lower dimensional theory only the sum of electric and Chern-Simons current is conserved.
	
	This also explains the absence of confinement seen in the lower dimensional theory \cite{Kogan:1995ei}: The monopoles on one side of the brane cannot tunnel to the other side, since on the other side, they are charged dyons, which can not be absorbed by the condensate. So there is no plasma of magnetic charges on the brane and no confinement.
	
	\section{Chern-Simons coefficient quantization}
    Our explicit construction -- detailed in the appendix -- allows for domain walls with $\Delta \Theta$ a multiple of $2\pi$.  This is compatible with the famous theorem by Vafa and Witten \cite{Vafa:1984xg}, which restricts the possible minima of the vacuum energy to occur for $\theta \in 2\pi \mathbb{Z}$. This reproduces the same quantization condition obtained from breaking 2+1d SU(2) Chern-Simons theory to U(1)$_\mathrm{c}$ \cite{Deser:1982vy}.
    
    The model described so far gives a nice understanding, of why magnetic events in the 2+1d theory lead to remnant electric charges. These electric charges represent the throats of electric flux tubes. We can also give a construction where the remnant charges have a simple description in the 2+1d effective theory. Let us add a doublet of fermions to the theory. Because of their color charge, they are confined in the bulk and we expect a localized fermion mode on the brane. As is well known, the monopoles acquire a fermion zero mode and monopoles with half integer charge appear in the spectrum. Flux tubes on the other hand are unstable to Schwinger-type pair creation in the extended theory. When a monopole crosses the brane in the extended theory --- instead of making a flux tube --- it's baryon number changes by one unit and a localized fermion of opposite baryon number is left on the brane.
	
	Notice that from the point of view of Yang-Mills theory we get a different periodicity in $\theta$ due to the axial anomaly. Therefore the Vafa-Witten theorem implies a different quantization condition for the Chern-Simons coefficient, namely $8\pi k\in\mathbb{Z}$. This is still consistent with the quantization condition obtained in \cite{Bachas:2009ve} from pure MCS theory. It does however allow values for $k$ that do not obey the quantization condition obtained by putting the 2+1d SU(2) on a 3-sphere and considering large gauge transformations. It is important to note, however that in our localization model, a large gauge transformation on the spatial sphere changes the number of instantons in the enclosed bulk. Due to the axial anomaly an instanton produces a fermionic zero mode and therefore does not leave the ground state invariant.
  
  	\section{Etc.}
	We can also try to understand what happens in the context of multiple branes, as those, when brought on top of each other, merge into a single brane. In other words from the view point of the lower dimensional EFT the $U(1)\times U(1)$ symmetry gets broken to a single diagonal $U(1)$ as the branes come on top of each other, as was first proposed in \cite{ArkaniHamed:1998ww}. In our context it is evident that one ends up with a jump in the axion field that is the sum of the jumps in the original branes, i.e. as two branes are brought on top of each other their CS levels add up. 
  
	Now we take a look at two monopoles of unit charge. Moving them across a brane of $\Dt=\pi$, each produces a fermion of charge $1/2$ on the brane. When we now exchange the fermions on the brane, the wavefunction picks up an additional anyonic phase
\be
	\delta=\pi.
\ee
Therefore the magnetic event remnant fermions actually behave like bosons. This is to be expected, because the interchange of monopoles before moving them across the brane would not have caused a nontrivial phase either.
    
    \section{Conclusions and Outlook}
    We have found an explicit mechanism to localize a MCS to a field-theoretic brane in 3+1 dimensions.
    Building on the Dvali-Shifman gauge field localization mechanism, we constructed a model in which the localization happens on an axionic domain wall. This naturally leads to a CS term in the 2+1d low energy theory.
    
    As expected our low energy theory exhibits the features deduced from classical MCS, namely a photon mass and the appearance of magnetic flux in the presence of electric charge. In terms of bulk physics those features can be understood as modified boundary conditions in Maxwell's equations due to the varying $\theta$-term across the bulk-brane interface.
    
    Quite generally for MCS theory obtained from some compactification, we have found that for large CS level the massive Kaluza-Klein-tower states can become lighter than the zero mode. I.e. for large CS level the effective theory behaves vastly different than one would expect from just a single gauge field with CS coupling.  
 
	Even more interestingly we have developed a nice understanding of the disappearance of confinement and the properties of magnetic events. Also the discreteness of the Chern--Simons coupling is related in a surprising way to bulk axion physics.    

	This work opens up several possible directions for future research, the most obvious being a generalization to more complicated gauge groups in the bulk and on the brane. Other interesting questions concern supersymmetric generalizations of our model with connections to string theory. In the original SQCD model given by Dvali-Shifman \cite{Dvali:1996xe}, the gluino-condensate order parameter $\langle\lambda\lambda\rangle$ plays the role of the axion. This implies that the conclusions drawn in our paper should also apply to the above mentioned field theoretic branes found in SQCD. This observation can shed some new light on the conjecture \cite{Witten:1997ep} that in the large N limit  the SQCD domain walls should assume a role analogous to D-branes in QCD string theory.
Using string dualities the effective world volume theory has already been conjectured to be a Chern--Simons gauge theory \cite{Acharya:2001dz}. It might also be worth exploring a possible realization of our setup in Seiberg-Witten theory \cite{Seiberg:1994rs}, where it could bear a relation to the phenomenon of wall-crossings. It would also be interesting to check whether a dual of our setup can lead to non-trivial applications in condensed matter systems.

\begin{acknowledgements}
We would like to thank Claudio Bunster, whose seminar talk inspired the present work. Furthermore we thank Gia Dvali, Andrei Khmelnitskiy, Nico Wintergerst and Anthony Zee for valuable discussions. 
The work of AP is supported by the Alexander von Humboldt foundation.
\end{acknowledgements}

	\appendix
	
	\section{Explicit potential} 
	Here we propose an explicit potential that can be used to realize the localization of Chern-Simons theory as described above. The potential for the Axion $\Theta(x)$ and Higgs $\phi(x)$ fields is
\begin{equation}
	V = A \cos(\Theta) + \left( B + C \cos(\Theta) \right) (\phi^a)^2 + \lambda (\phi^a)^4 + \dotsb
\end{equation}
As long as $|B| < C$ and $A$ and $C$ are of the same sign, the potential has global minima where $\theta$ is a multiple of $2\pi$ and $\phi = 0$ in this theory. This corresponds to the confining vacuum in the bulk. Clearly, axionic domain walls exist and near the center of such a wall, the Higgs field must take a nonzero vev to minimize the potential. An exemplary field configuration for such a domain wall is shown in fig. \ref{fig brane field} (left). Terms of higher order in $\phi$ or $\cos(\Theta)$ will not change the story much as long as their coefficients are sufficiently small. Note however that $B-C$ must be sufficiently negative to allow for a Higgs-mode to condense.

	\section{Image charge details}
	Here we will give some details of our calculation of the electromagnetic field sourced by a point charge on the domain wall. Consider the simplified situation described in the text, where the axion vev does not change continuously by $\Dt$, but in two discrete jumps at the left and right boundary of the U(1) domain. For the purpose of determining the correct boundary conditions, the dual superconductor in the bulk can be modeled by a medium of singular permeability/permittivity. For one charge on the brane, an infinite series of image charges is recursively built up. For a given electric and magnetic charge, the charge-vector of its mirror image is determined by the following matrix:
	\begin{equation}
	\left(
\begin{array}{cc}
 \frac{-\theta _{\mathrm{dsc}}^2+\theta _{\mathrm{vac}}^2+\pi ^2}{\left(\theta _{\mathrm{vac}}-\theta _{\mathrm{dsc}}\right){}^2+\pi ^2} & \frac{\theta _{\mathrm{dsc}} \left(\theta _{\mathrm{vac}} \left(\theta _{\mathrm{vac}}-\theta _{\mathrm{dsc}}\right)+\pi ^2\right)}{\pi  \left(\left(\theta _{\mathrm{vac}}-\theta _{\mathrm{dsc}}\right){}^2+\pi ^2\right)} \\
 \frac{4 \pi  \left(\theta _{\mathrm{dsc}}-\theta _{\mathrm{vac}}\right)}{\left(\theta _{\mathrm{vac}}-\theta _{\mathrm{dsc}}\right){}^2+\pi ^2} & -\frac{-\theta _{\mathrm{dsc}}^2+\theta _{\mathrm{vac}}^2+\pi ^2}{\left(\theta _{\mathrm{vac}}-\theta _{\mathrm{dsc}}\right){}^2+\pi ^2}
\end{array}
\right)
	\end{equation}
	where $\theta_\mathrm{vac}$ is the axion-vev in the U(1) vacuum on the brane and $\theta_\mathrm{dsc}$ is the axion-vev in the dual-superconductor on the other side of the respective boundary. 
	
	For the simplest case, when $\Dt=2\pi$ is symmetrically distributed and $\theta_\mathrm{vac}=0$, the mirror charges contributing to the electric field at the center of the brane are just an evenly spaced sequence of alternating sign. After Fourier expanding the charge density, the field created by the Fourier mode $\rho_k \propto \cos((2k+1) \pi z / (2 L))$ can be evaluated exactly. At asymptotically large distance from the charges, the $k=0$ mode dominates and reproduces the expected Yukawa form
	\begin{equation}
		|E(r,\theta,z=0)| \propto \left( \frac{1}{\sqrt{r}} + \mathcal{O}(1/r^{3/2}) \right) \exp\left(-\frac{\pi r}{2 L}\right)
	\end{equation}
	
\begin{figure}
		\includegraphics{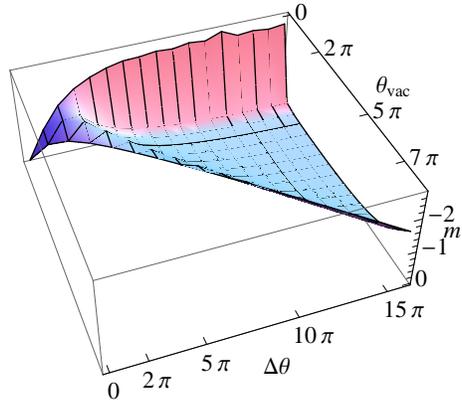}
		\caption{\label{fig mass}Effective mass of the lightest mode seen by a brane observer as a function of the discontinuities in $\theta$}
	\end{figure}

	For generic $\Dt$ and when $\Dt$ is distributed unevenly to the left and right boundary, the effective mass of the lightest mode can be evaluated numerically. The results are displayed in \ref{fig mass}. As expected, for a small values of $\Dt$ we see a linear growth of the mass with $\Dt$. This is precisely the Chern-Simons mass of the previously-massless 2+1d photon. In the case of larger $\Dt$ however, the situation is more complicated. When $\Dt$ is distributed unevenly, we see a falloff in the effective mass proportional to $\Dt^{-1}$. This can be interpreted as one of the two dof of a massive KK-mode becoming light, as remarked in the text. The evenly distributed case however is peculiar, as the effective mass saturates for large $\Dt$ at the value of the mass of the lightest KK-mode in the absence of a Chern-Simons term. In this case there is an enhanced symmetry, implemented by first performing a reflection about the brane and then inverting the axion field. It implies that both dof in each KK level have to stay degenerate.
		
\end{document}